\documentstyle[prl,aps,preprint]{revtex}

\def\upleftarrow#1{\overleftarrow{#1}}
\def\thru#1{\mathrel{\mathop{#1\!\!\!/}}}

\begin{document}
\tighten

\title{BREAKUP OF HADRON MASSES AND \\
ENERGY-MOMENTUM TENSOR OF QCD\thanks {%
This work is supported in part by funds provided by the
U.S.  Department of Energy (D.O.E.) under cooperative agreement
\#DF-FC02-94ER40818.}}

\author{Xiangdong Ji}

\address{Center for Theoretical Physics \\
Laboratory for Nuclear Science \\
and Department of Physics \\
Massachusetts Institute of Technology \\
Cambridge, Massachusetts 02139 \\
{~}}

\date{MIT-CTP-2407 ~~~ Submitted to: {\it Phys.~Rev.~D\/} ~~~ January 1995}

\maketitle

\begin{abstract}
Hadron masses are shown to be separable in QCD into contributions
of quark and gluon kinetic and potential energies, quark
masses, and the trace anomaly. The separation is based on
a study of the structure of the QCD energy-momentum tensor
and its matrix elements in hadron states. The paper contains
two parts. In the first part, a detailed discussion of the
renormalization properties of the energy-momentum tensor
is given.  In the second part, a mass separation formula
is derived and then applied to the nucleon, pion, and the QCD vacuum.
Implications of the results on hadron structure and
non-perturbative QCD dynamics are discussed.
\end{abstract}
\pacs{xxxxxx}

\narrowtext

\section{Introduction}

For any physical system, a good knowledge about its mass structure is
helpful in understanding the underlying dynamics of the
system. The ground state energy of nuclei as a function of the mass
number reveals the nuclear shell structure and the saturation property
of nuclear force~\cite{PB}. An organized pattern of hadron masses led
to the speculation of the quark substructure of
hadrons~\cite{GN}. Likewise, any knowledge about structure of
hadron masses in terms of its underlying constituents, quarks and
gluons, can be useful to unlocking the physics of Quantum
Chromodynamics (QCD) in strong coupling region.

At classical level, the lagrangian of Chromodynamics
is invariant under scale transformation if quark masses are
neglected, and thus hadron masses necessarily vanish.
Radiative quantum effects break the scale symmetry~\cite{CCJ} and
introduce a dimensional parameter
$\Lambda_{\rm QCD}$ through dimensional transmutation~\cite{CW}.
[For a classical discussion on the
scale anomaly, see Ref.~\cite{CCJ}]. Although $\Lambda_{\rm QCD}$
is well-determined through scale-breaking effects in
high-energy processes~\cite{PDT}, the physical
mechanism for generating the scale at low energy is
not quite clear. [It is encouraging, though, that
the scale determined from hadron spectrum using lattice QCD
is consistent with that determined at
high energy~\cite{CF}.]

In the past, scale generation at low energy is largely understood
in two seemingly-unrelated pictures, and so are masses of
hadrons. The first picture emphasizes the aspect of color confinement,
through which quarks in hadrons are confined to a cavity of
radius $\sim$ 1 fm. A simple, representative model is the MIT bag, in
which the mass of the nucleon is generated from the quark kinetic energy
and vacuum pressure~\cite{BAG}. A dimensional analysis shows that
the quark kinetic energy accounts for 3/4 of the mass and the
vacuum pressure accounts for the rest 1/4. In some loose sense, the
vacuum pressure models effects of long-wavelength gluons.  The
second picture for scale generation emphasizes chiral symmetry
breaking. Through such phase transition in the QCD vacuum, light
quarks acquire a constituent mass of order 300 MeV\@.  Masses of
hadrons are then approximately the sum of the constituent masses
and quark kinetic and potential energies~\cite{IK}. Both pictures work
quite well for hadron spectra and other physical observables.  It is
not known, however, which picture or which combination of the
two is closer to reality.

In a Letter article published recently~\cite{JI}, this author showed
that an insight about the mass structure of hadrons can be obtained
through a study of the energy-momentum tensor of QCD\@.  The result is
a separation of hadron masses into contributions from quark and gluon
kinetic and potential energies, the current quark masses, and the
trace anomaly. The last part is a direct result of scale
symmetry breaking, and is analogous to the vacuum pressure empirically
introduced in the MIT bag model. Though the scale of other
contributions is determined by the anomaly, relative magnitudes
reflect important aspects of the low-energy quark-gluon dynamics. The
separation here is analogous to the {\it virial theorem} for a simple
harmonic oscillator and the hydrogen atom.

The goal of this paper is to provide more field theoretical basis for
the mass separation and to extend its application to other hadrons
besides the nucleon. The field theoretical discussion mainly answers
questions like: Why the QCD hamiltonian is finite? Why it can be
separated into gauge-invariant and finite pieces? How the
renormalization scale dependence affects the separation? Why there is
an extra term in the hamiltonian from the trace anomaly? etc. An
application of the mass separation to a specific hadron requires
knowledge of two matrix elements: the momentum fraction of the hadron
carried by quarks in the infinite momentum frame and the quark scalar
charges.  The matrix elements are known to a good accuracy in the
nucleon and pion. However, there is no firm
estimate of these for other hadrons. Nonetheless, a plausible assumption
leads to a crude picture for general mass partition.

The paper consists of two main sections. In Section~II, we start with
the separation of the QCD energy-momentum tensor into the traceless
and trace parts. Then we consider renormalization of both parts in
dimensional regularization and the covariant gauge fixing.  We present
a derivation the trace anomaly from the point of view of
operator mixing. We also discuss structure of the mixing matrix for the
traceless part of the energy-momentum tensor.  As a by-product,
we derive the dilatation current in QCD and study
its anomalous Ward identities (Callan-Symanzik equation). For a
phenomenology-oriented reader, this section can be skipped. In
Section~III, we study the hadron matrix element of the
tensor and derive a separation of the QCD hamiltonian and hence
hadron masses. For the application to the nucleon, we focus on discussing
physical significance of the different contributions.
The result for the pion strongly supports the concept that the pion is
a collective vacuum excitation. Finally, comments are made about
general features of the mass separation for other hadrons and the QCD
vacuum.

\section{Renormalization of The QCD Energy-Momentum Tensor }

In this section, we discuss renormalization properties of the QCD
energy-momentum tensor. A similar but less extensive discussion was
first made by Freedman, Muzinich, and Weinberg~\cite{FMW} for
non-Abelian gauge theory. Since then, many papers relevant to the
subject have appeared in the literature.  The present discussion is
not a review of the subject. In particular, there are still open
questions that are under current debate. Rather, we will focus on the
following three aspects that are most useful for our discussion of
hadron masses.  First, the energy-momentum tensor can be uniquely
separated into the trace and traceless parts which are to be
separately renormalized.  Second, the trace part of the tensor
is shown to be finite according to that the renormalized
Green's functions are finite
functions of renormalized masses and couplings. The result is
equivalent to a new derivation of the trace anomaly originally given
in Refs.~\cite{CDJ,NIL}.  Third, the traceless part of the tensor is
shown to be finite using Ward identities related to space-time
translational symmetry.  A study of scale symmetry and related Ward
identities (Callan-Symanzik equation) is included in the end of this
section as a by-product of the previous discussion.

The fundamental QCD lagrangian reads,
\begin{equation}
      {\cal L}_{\rm QCD}= \bar \psi (i\thru D - m)\psi - {1\over 4}
         F^{\mu\nu }F_{\mu\nu } \; ,
\label{lag}
\end{equation}
where $\psi$ is quark fields, carrying implicit flavor, color and
Dirac indices, $D^{\mu} = \partial^\mu+igA^{\mu}$ is a covariant
derivative with $A^\mu=A^{\mu a}t^a$ being the gauge potential (Tr$\
t^at^b=\delta^{ab}/2$), $m $ is the quark mass matrix in flavor
space. The gluon field strength has the usual non-Abelian expression,
\begin{equation}
       F^{\mu\nu a} = \partial^\mu A^{\nu a}
            - \partial^\nu A^{\mu a} - gf^{abc}A^{\mu b}A^{\nu c} \; ,
\end{equation}
where $f^{abc}$ is the structure constant of the gauge group. To simplify
notations, we shall omit color indices if no confusion
arises.  All quantities without further specification are bare ones.

To discuss renormalization, we work in
dimensional regularization and (modified) minimal subtraction
scheme. An important virtue of this scheme is that it does not modify
the basic QCD lagrangian except all Lorentz indices and space-time
integrations are taken to be $d$ dimensional. [The Pauli-Villars
regularization, some momentum cut-off schemes, or lattice cut-off all
modify the lagrangian in a more significant way.]  We use perturbation
theory to explore large momentum behavior, and thus a gauge fixing is
needed.  We choose the usual prescription for use of the covariant
gauge, which introduces two extra terms in the lagrangian: the gauge
fixing term,
\begin{equation}
      { \cal L}_{\rm gf} = -{1\over 2\xi} (\partial^\mu A_\mu)^2  \;  ,
\end{equation}
and the Faddeev-Popov ghost term,
\begin{equation}
      {\cal L}_{\rm gs} = \partial^\mu \bar \omega D_\mu \omega \;  .
\end{equation}
Therefore the total effective lagrangian,
\begin{equation}
       {\cal L}_{\rm eff} = {\cal L}_{\rm QCD}
       + {\cal L}_{\rm gf} + {\cal L}_{\rm gs}  \;  ,
\end{equation}
enters the rest discussion.

The energy-momentum tensor can be derived from the fact that the
action $S=\int d^dx {\cal L}_{\rm eff}$ is invariant under space-time
translation. The result is well-known~\cite{FMW},
\begin{eqnarray}
      T^{\mu\nu} &=& - g^{\mu\nu} {\cal L}_{\rm eff}
           - F^{\mu\alpha }F^{\nu }_{\ \alpha}
            + {1\over 2}\bar \psi iD^{(\mu} \gamma^{\nu)} \psi
           +  {1\over 2}\bar \psi i \upleftarrow{D}^{(\mu} \gamma^{\nu)} \psi
           \nonumber \\
&&\quad {} -  g^{\mu\nu} \xi^{-1} \partial^\alpha(A_\alpha
             \partial \cdot A)
               +  2\xi^{-1} A^{(\mu }\partial^{\nu)} (\partial \cdot A)
              \nonumber \\
&&\quad {} + 2\partial^{(\mu} \bar \omega D^{\nu)} \omega
           - {\delta S \over \delta A_\mu}A_\nu
        - {\delta S \over \delta \psi} {1\over 8}[\gamma^\mu,\gamma^\nu]\psi
      - \bar \psi {1\over 8}[\gamma^\mu, \gamma^\nu]{\delta S\over \delta\bar
     \psi} \;  .
\label{tmunu}
\end{eqnarray}
where $(\mu\nu)$ means symmetrization of the indices and $\upleftarrow
{D}^\mu = - \upleftarrow {\partial}^\mu + igA^\mu$. There are two
standard methods to derive the above expression. First, one can write
down the canonical Noether current associated with space-time
translational symmetry and then improves it with the Belinfante
procedure~\cite{JAC}.  Second, one can follow Ref.~\cite{FMW},
deriving a current coupling to gravity. Both methods yield the same
tensor for QCD, though not for the $\phi^4$ theory.

The last three terms in (\ref{tmunu}) are usually ignored in other
references because they vanish when the equations of motion for gluon
and quark fields are used. When they are inserted into a Green's
function, their role is to change polarizations of external lines and
thus they are finite operators. When they are included in the tensor,
the Ward identities related to space-time translation symmetry
take a simpler form~\cite{NIL}.  Without these terms, the energy
momentum tensor is symmetric,
\begin{equation}
      T^{\mu\nu} = T^{\nu\mu} \;  .
\end{equation}
In the following discussion, we concentrate on the
symmetric part of the tensor only.

The symmetric part of the energy-momentum tensor can
be split into a sum of traceless and trace parts.
Although a splitting like this is not unique in general, we define
the splitting in the following unambiguous way,
\begin{equation}
               T^{\mu\nu} = \bar T^{\mu\nu} + \hat T^{\mu\nu}  \;  ,
\end{equation}
where the traceless ($\bar T^{\mu\nu}$) and trace parts
($\hat T^{\mu\nu}$) are,
\begin{eqnarray}
         \bar T^{\mu\nu} &=& T^{\mu\nu} -
    {1\over d}g^{\mu\nu}T^\alpha_{\ \alpha}  \;  ,
        \nonumber \\
\hat T^{\mu\nu} &=&  {1\over d} g^{\mu\nu}T^\alpha_{\ \alpha} \;  .
\end{eqnarray}
Notice that we have defined the trace consistent with dimensional
regularization, under which tensors of different ranks do not
mix. Therefore the necessary and sufficient condition for the
energy-momentum tensor to be finite is that the traceless and trace
parts are separately finite. This is indeed the case as we shall see
below.

\subsection{The Trace Part of the Energy-Momentum Tensor}

Let us first consider the trace-part of the energy momentum tensor,
\begin{eqnarray}
T^\alpha_{\ \alpha} &=&\bar \psi m \psi+\epsilon
      \left[ O_F+\bar O_{\rm gf} \right] - (2-\epsilon)
          \partial^\alpha \left[ \xi^{-1}A_\alpha\partial \cdot A
           + \bar \omega D_\alpha \omega \right]
           \nonumber \\
&& \quad {} - (3-\epsilon)\bar \psi {\delta S \over \delta \bar \psi}
            - (2-\epsilon) \bar \omega {\delta S \over \delta \bar \omega}
            - \left( 1-{\epsilon\over 2} \right)
           A_\mu {\delta S \over \delta A_\mu}  \;  ,
\label{trace}
\end{eqnarray}
where we have defined,
\begin{eqnarray}
O_F &=&  -{1\over 4}F^{\alpha\beta}F_{\alpha\beta}  \;  ,
           \nonumber \\
O_{\rm gf} &=&  - (2\xi)^{-1}(\partial\cdot A)^2  \;  ,
           \nonumber \\
\bar O_{\rm gf}  & =&  O_{\rm gf} - {1\over 2} A_\mu
           {\delta S\over\delta A_\mu}\;  .
\end{eqnarray}
In Refs.~\cite{CDJ,NIL}, it was shown that the trace can be expressed
in terms of a linear combination of a set of renormalized operators
with finite coefficients.  Here we would like to derive the same
result from the operator mixing point of view~\cite{Brown}, by
extending the discussion in Ref.~\cite{KS}.

First consider the equations-of-motion-related operators,
\begin{eqnarray}
O_A  & =&  A_\mu {\delta S \over \delta A_\mu}  \;  ,
           \nonumber \\
O_\psi &=&  \bar \psi {\delta S \over \delta \bar \psi} = \bar \psi
           (i\thru D-m)\psi   \;  ,
           \nonumber \\
O_\omega & =&  \bar \omega {\delta S \over \delta \bar \omega}= \bar
           \omega\partial^\alpha D_\alpha \omega \;  .
\end{eqnarray}
A zero momentum insertion of these operators into a renormalized
Green's function produces the number of external lines. Therefore all
these operators are finite and need no renormalization.

The total derivative operator in the third term in Eq.~(\ref{trace}) is a
BRST variation of $
\partial_\alpha(A^\alpha\bar \omega)$, i.e.,
\begin{equation}
           \partial^\alpha\delta_{\rm BRST}
        \left( A_\alpha\bar \omega \right)
           = \partial^\alpha
           \left[ \xi^{-1} A_\alpha\partial \cdot A
           + \bar \omega D_\alpha \omega \right] \;  .
\end{equation}
As such it has vanishing physical matrix elements at non-zero
momentum~\cite{JL,COL}. At zero momentum, the operator vanishes identically
as it contains no poles. Since other dimension-4
operators do not mix with it, it is a finite operator by itself.

Renormalization of the rest three operators can be worked out
by studying the renormalized Green's function generating functional,
\begin{eqnarray}
&& Z(J,\eta,\bar\eta, \chi,\bar\chi)
           \nonumber \\
&&\qquad   = { \int  D \left( A^R,\psi^R,\bar\psi^R,
                      \omega^R,\bar\omega^R \right)
                \exp{i \left[ S + \int (J^\mu A_\mu^R +\bar \eta \psi^R
                +\bar \psi^R \eta + \bar \chi \omega^R
                      + \bar \omega^R \chi) \right]}
       \over \int D \left( A^R,\psi^R,\bar\psi^R, \omega^R,\bar\omega^R \right)
                  \exp{iS} }  \;  ,
\end{eqnarray}
where the renormalized fields are related to bare fields by
the multiplicative renormalization constants,
\begin{equation}
   \psi = Z_2 \psi^R  \;  ,
           \qquad    \omega = \tilde Z \omega^R  \;  ,
           \qquad    A_\mu = Z_3 A^R_\mu  \;  .
\end{equation}
Since $ Z(J,\eta,\bar\eta, \chi,\bar\chi)$ is finite, its derivatives
with respect to the renormalized quark masses $m^R$, the gauge
coupling constant $g^R$, and the gauge fixing parameter $\xi^R$ are
also finite. In dimensional regularization and minimal subtraction
scheme, all the renormalization constants are independent of quark
masses and the renormalization factors for the gauge coupling $Z_g$
and the quark masses $Z_m$ are independent of the gauge-fixing
parameter~\cite{MUTA}.  According to these, we find the following
quantities are finite,
\begin{eqnarray}
O_m  & =&  \bar  \psi m \psi   \;  ,
           \nonumber \\
\bar O_{\rm gf}^R &=&
           \left(1+\xi^R {\partial \ln Z_3 \over \partial \xi^R} \right)
           \bar O_{\rm gf}
           -\xi^R {\partial \ln Z_2 \over \partial \xi^R} O_\psi
           -\xi^R {\partial \ln \tilde Z \over \partial \xi^R} O_\omega  \;  ,
           \nonumber \\
O_F^R &=&
           \left( 1+g^R{\partial \ln Z_g \over \partial g^R} \right)
           O_F +
           \left( g^R{\partial \ln Z_g \over \partial g^R} -
           \xi^R{\partial \ln Z_3 \over \partial \xi^R} +{1\over 2}
           g^R{\partial \ln Z_3 \over \partial g^R} \right)
           \bar O_{\rm gf}
           \nonumber \\
&& \quad {} +  \left( \xi^R {\partial \ln Z_2 \over \partial \xi^R} -{1\over 2}
           g^R{\partial \ln Z_2 \over \partial g^R} \right) O_{\psi}
           + \left( \xi^R {\partial \ln \tilde Z \over \partial \xi^R}
                      -{1\over 2}
           g^R{\partial \ln \tilde Z \over \partial g^R} \right) O_{\omega}
           \nonumber \\
&& \quad {}  + {1\over 2}g^R{\partial \ln Z_m \over \partial g^R} O_m  \;  .
\end{eqnarray}
{}From the above equations, we can solve $O_F$ and $\bar O_{\rm gf}$
in terms of the renormalized ones,
\begin{equation}
    \epsilon \left( O_F + \bar O_{\rm gf} \right)
      = \left( -2\beta/g^R \right)  O_F^R
           + \left( -2\beta/g^R+\gamma_3 \right)
      \bar O^R_{\rm gf} - \tilde \gamma O_\omega - \gamma_2 O_\psi
          + \gamma_m O_m \;  ,
\label{reno}
\end{equation}
where we have defined the anomalous dimensions,
\begin{eqnarray}
\gamma_{2,3,m} &=&  \mu {d\ln Z_{2,3,m}\over d\mu} \;  ,
           \nonumber \\
\tilde \gamma &=&  \mu {d\ln \tilde Z\over d\mu} \;  ,
           \nonumber \\
\beta &=&  -g^R\mu {d\ln Z_g\over d\mu}  \; .
\end{eqnarray}
Since all these quantities have no $1/\epsilon$ poles, the right hand side of
Eq.~(\ref{reno}) is finite.

Inserting Eq.~(\ref{reno}) into Eq.~(\ref{trace}),
we find the trace part of the
energy-momentum tensor expressed in terms of
finite, renormalized operators,
\begin{eqnarray}
T^{\alpha}_{\ \alpha }  &=&  \left( -2\beta/g^R \right)
           \left( O_F^R+\bar O_{\rm gf}^R \right)
           + \left( 1+\gamma_m \right) O_m
           -2 \partial^\alpha \delta_{\rm BRST}
           \left( A_\alpha\omega \right)
           \nonumber \\
&& \quad {} + \gamma_3 O_{\rm gf}^R
           - \left( 1+{\gamma_3\over 2} \right) O_A
           - \left( 2+\tilde \gamma \right) O_\omega
           - \left( 3+\gamma_2 \right) O_\psi \;  .
\end{eqnarray}
Two comments are in order at this point. First,
composite operators defined through path-integral formalism
have implied subtraction of their vacuum expectations.
Thus they are said ``normal ordered,'' but not in the usual
sense of relative to perturbative vacuum. Second, although
individual term in the above equation depends on the
renormalization scale ($\mu^2$), but the sum does not.

According to the Joglekar and Lee theorems~\cite{JL,COL},
the matrix elements of
$O_A$, $O_\psi$, and $O_\omega$, and BRST exact operators
vanish in a physical state. Thus as far as hadron matrix elements
are concerned, we effectively have,
\begin{equation}
       T^{\alpha}_{\ \alpha } = \left( -2\beta/g^R \right) O_F^R
           + \left( 1+\gamma_m \right) O_m \;  .
\end{equation}
This suggests we write the trace part of the energy-momentum tensor
as,
\begin{equation}
      \hat T^{\mu\nu} = \hat T^{\mu\nu}_a \left( \mu^2 \right)
           + \hat T^{\mu\nu}_m \left( \mu^2 \right )  \;  ,
\end{equation}
where $T^{\mu\nu}_a=- \left( 2\beta/g^R \right) O_F^R
           \left( g^{\mu\nu}/4 \right)$ and
$T^{\mu\nu}_m = \left(1+\gamma_m \right) O_m \left( g^{\mu\nu}/4 \right)$.

\subsection{The Traceless Part of the Energy-Momentum Tensor}

The traceless part of the energy momentum tensor can be written
as a sum of four parts,
\begin{equation}
            \bar T^{\mu\nu}
               = \bar T^{\mu\nu}_g + \bar T^{\mu\nu}_q
               + \bar T^{\mu\nu}_{\rm gv} + E^{\mu\nu} \;  .
\end{equation}
The gauge-invariant gluon part $\bar T^{\mu\nu}_g$ is,
\begin{equation}
                T^{\mu\nu}_g = -F^{(\mu\alpha}F^{\nu)}_{\ \alpha} \;  ,
\end{equation}
where and henceforth the symbol $(\mu\nu)$ also means
subtraction of the trace.
The gauge-invariant quark part is
\begin{equation}
             T^{\mu\nu}_q  = {1\over 2}\bar \psi
             iD^{(\mu} \gamma^{\nu)} \psi + {1\over 2}\bar \psi
          i\upleftarrow{D}^{(\mu} \gamma^{\nu)} \psi \;  .
\end{equation}
The gauge variant part is,
\begin{equation}
            T^{\mu\nu}_{\rm gv} = \xi^{-1}A^{(\mu}\partial^{\nu)}
              \partial\cdot A
              + \partial^{(\mu}\bar \omega D^{\nu)} \omega   \;  ,
\end{equation}
which is the BRST variation of $\partial^{(\mu }\bar \omega A^{\nu)}$.
Finally, the gluon-equation-of-motion-related operator is,
\begin{equation}
       E^{\mu\nu} = - A^{(\mu}{\delta S \over \delta A_{\nu)}}  \;  ,
\end{equation}
which is finite.

According to the Joglekar and Lee theorems~\cite{JL}, the above
four operators close under
renormalization. Furthermore, the mixing matrix takes the following
form,
\begin{equation}
\pmatrix{
           T^{\mu\nu}_q \cr
           T^{\mu\nu}_g \cr
           T^{\mu\nu}_{\rm gv} \cr
           E^{\mu\nu}  }
= \pmatrix{
           Z_{qq} & Z_{qg} & Z_{qa} & Z_{qe} \cr
           Z_{gq} & Z_{gg} & Z_{ga} & Z_{ge} \cr
           0      &  0      &  Z_{aa} & Z_{ae} \cr
           0      &  0      &  0      &  1      }
\pmatrix {
           T^{\mu\nu}_q \cr
           T^{\mu\nu}_g \cr
           T^{\mu\nu}_{\rm gv} \cr
           E^{\mu\nu}
}^R  \;  .
\label{mix}
\end{equation}
That is, the gauge variant operators do not mix with gauge invariant
operators and the equations of motion operators do not mix with
any other operators.

To find the $Z$'s, one needs to study
Green's functions with one insertion of the composite operators
in perturbation theory. However, we can find constraints
among the renormalization constants by studying
Ward identities related to space-time translational symmetry.
Calculating the divergence of the energy-momentum tensor without
using the equations of motion, we find,
\begin{equation}
   \partial_\mu T^{\mu\nu} = -{\delta S \over \delta A_\mu} \partial^\nu
           A^\mu -{\delta S \over \delta \psi} \partial^\nu
            \psi - {\delta S \over \delta \bar \psi} \partial^\nu
            \bar \psi- {\delta S \over \delta \omega } \partial^\nu
           \omega - {\delta S \over \delta \bar \omega} \partial^\nu
           \bar \omega \;  .
\end{equation}
When the above equation is inserted into a renormalized
Green's function (Ward identity), the right-hand side
simply replaces the elementary fields by their derivatives,
and thus is finite. Hence the divergence of the energy
momentum tensor is a finite operator. The only remaining
counter term for the tensor itself is of form
\begin{equation}
      \left( \partial^\mu\partial^\nu - g^{\mu\nu}\partial^2 \right)
           f \left( A,\omega,\bar\omega \right)  \;  ,
\end{equation}
where $f$ is a dimension-two function of its arguments.
However, since the trace part of the energy-momentum
tensor is finite, Lorentz symmetry allows only counter-term
tensors that are symmetric and traceless. Thus the traceless
part of the tensor must be finite by itself~\cite{NIL}.

The finite traceless part of the energy-momentum tensor
imposes the following constraints on the renormalization constants
in minimal subtraction scheme,
\begin{eqnarray}
           Z_{qq} + Z_{gq}   &=&  1  \;  ,   \nonumber \\
           Z_{qg} + Z_{gg}   &=&  1  \;  ,    \nonumber \\
           Z_{qa} + Z_{ga} + Z_{aa}  &=&  1  \;  ,  \nonumber \\
           Z_{qe} + Z_{ge} + Z_{ae}  &=&  0  \;  .
\end{eqnarray}
Thus in the scheme, we have,
\begin{equation}
            T^{\mu\nu}_q + T^{\mu\nu}_g +
             T^{\mu\nu}_{\rm gv} + E^{\mu\nu} =
           T^{\mu\nu R}_q \left( \mu^2 \right)
           + T^{\mu\nu R}_g \left( \mu^2 \right) +
             T^{\mu\nu R}_{\rm gv}
           \left( \mu^2 \right)
           + E^{\mu\nu}\;  .
\end{equation}
Although individual term on the right hand side depends on
the renormalization scale, the sum does not.
Again, according the Joglekar-Lee theorems,
$T^{\mu\nu R}_{\rm gv}(\mu^2)$ and $E^{\mu\nu}$ have vanishing
physical matrix elements, and for practical purposes, we
can keep only the gauge-invariant quark and gluon contributions
in the traceless part of the energy-momentum tensor,
\begin{equation}
      \bar  T^{\mu\nu} = \bar T^{\mu\nu R}_q(\mu^2)
        + \bar T^{\mu\nu R}_g(\mu^2) \;  .
\end{equation}

Recently, some questions arise in the literature
about validity of the Joglekar and Lee theorems~\cite{HV,CS},
in particular, regarding
the form of the mixing matrix appearing in Eq.~(\ref{mix}).
Harris and Smith \cite{HS} pointed out that so long as
one is working with composite operators at non-zero
momentum transfer, all the theorems
should remain valid. Collins and Scalise~\cite{CS}, on the other hand,
have worked at zero momentum transfor. They showed that the
on-shell limit for gauge bosons is subtle and potentially
causes problems. However, it appears that if one works with
off-shell Green's functions or physical hadron states,
the operator mixing shall follow the standard theorems.
More work is certainly needed in this direction to clear up
the issue.

\subsection{Scale Symmetry and the Anomalous Ward Identities}

The above discussion on mixing of dimension-4 operators
and the trace of the energy momentum tensor permits a simple
explanation for the anomalous breaking of scale symmetry
and a derivation of Callan-Symanzik equation (or the anomalous
Ward identity). Consider the scale (dilatation) transformation
in $d$ dimension,
\begin{eqnarray}
        x & \to&  \lambda x    \;  , \nonumber \\
        \psi(x)  &\to&  \lambda^{d-1\over 2} \psi(\lambda x)  \;  ,
                      \nonumber \\
        A^\mu(x) &\to&  \lambda^{{d\over 2}-1}A^\mu(\lambda x)  \;  ,
                      \nonumber \\
        \omega(x) &\to&  \lambda^{{d\over 2}-1}\omega(\lambda x) \;  .
\end{eqnarray}
The change in the effective QCD lagrangian is a total derivative
plus symmetry breaking terms. There are two types of symmetry breaking
effects: quark masses and the gauge coupling, the latter has
dimension $\epsilon/2$ in dimensional regularization.
Thus,
\begin{equation}
        \delta {\cal L}_{\rm eff} = \partial^\mu(x_\mu {\cal L}_{\rm eff})
           + \bar \psi m \psi -{\epsilon\over 2} g{\partial {\cal L}_{\rm eff}
            \over \partial g} \;  .
\end{equation}
The last term is clearly not gauge invariant. A simple rearrangement
reveals that it consists of operators $O_F$ and $\bar O_{\rm gf}$ defined
in the previous section plus a total derivative term,
\begin{equation}
           - g{\partial {\cal L}_{\rm eff} \over \partial g}
           = \partial^\mu
           \left( F_{\mu\nu}A^\nu + \xi^{-1}(\partial\cdot A )A_\mu \right)
        +2O_F + 2\bar O_{\rm gf} \;  .
\end{equation}
Since $O_F$ and $\bar O_{\rm gf}$ diverges like $1/\epsilon$, the scale
symmetry is broken anomalously at the quantum level.

On the other hand, with use of the equations of motion, one has,
\begin{equation}
        \delta {\cal L}_{\rm eff}
                      = \partial^\mu\left({\partial {\cal L}_{\rm eff}
         \over \partial \partial_\mu \phi_i} \delta \phi_i \right)  \;  ,
\end{equation}
where $\phi_i$ is a generic notation for all the fields.
According to this, we can define a dilatation current corresponding
to the scale transformation,
\begin{equation}
J_D^\mu =  {\partial {\cal L}_{\rm eff} \over \partial \partial_\mu \phi_i}
           \delta \phi_i  - x^\mu {\cal L}_{\rm eff}
           - {\epsilon\over 2}
           \left( F^{\mu\nu}A_\nu + \xi^{-1}(\partial\cdot A)A^\mu \right)
                      \;  .
\end{equation}
The symmetry breaking terms now appear as divergence of the current,
\begin{equation}
        \partial_\mu J^\mu_D = \bar \psi m \psi
       + \epsilon \left( O_F + \bar O_{\rm gf} \right) \;  .
\end{equation}
To prove that, use of the equations of motion is essential.

The dilatation current has a simple relation with the energy-momentum
tensor in Belinfante's form. To see that, we use the definition of the
canonical energy-momentum tensor $T_C^{\mu\nu}$ to write,
\begin{equation}
        J_D^\mu =  T^{\mu\nu}_C x_\nu
           + {\partial {\cal L}_{\rm eff} \over \partial \partial_\mu
           \phi_i} d_{\phi_i}  {\epsilon\over 2}
           \left( F^{\mu\nu}A_\nu
                + \xi^{-1}(\partial\cdot A)A^\mu \right)  \;  ,
\end{equation}
where $d_\phi$ is the canonical dimension for $\phi_i$ field.
Using the Belinfante improvement, one has,
\begin{equation}
            J_D^{\mu} = T^{\mu\nu}x_\nu + {\partial {\cal L}_{\rm eff}
         \over \partial \partial_\mu \phi_i}
           \Big( \Sigma^{\mu\rho} + g^{\mu\rho} d_\phi \Big)_{ij}\phi_j
           - {\epsilon\over 2}
           \left( F_{\mu\nu}A^\nu
                + \xi^{-1}(\partial\cdot A)A_\mu \right)   \;  ,
\end{equation}
where the second term is called field-virial, which has a part that
cancels the last gauge non-invariant term,
\begin{equation}
           J_D^\mu =T^{\mu\nu}x_\nu
           + \left( 1-{\epsilon\over 2} \right)
           \Big( \partial^\mu \bar \omega \omega
               +\bar \omega D^\mu \omega \Big)
        +(2-\epsilon)\xi^{-1}(\partial\cdot A)A^\mu \;  .
\end{equation}
The remaining terms cannot be written as a total derivative. According
to Callan, Coleman, and Jackiw~\cite{CCJ}, this means that
conformal symmetry present in the basic QCD lagrangian
is broken by the gauge fixing. This fact is of course well-known
in perturbation theory where a gauge-fixed gluon propagator
does not have conformal symmetry. However, the breaking term is a
BRST-exact operator plus the ghost equation of motion and thus
has no physical consequences.

To derive the anomalous Ward identities associated with scale
transformation, we consider corresponding change in the Green's function
generating functional $Z(J_\mu,\eta,\bar\eta,\chi,\bar\chi)$.
A simple derivation yields the following equation,
\begin{equation}
         \sum_i \left( d_{\phi_i} + x_i{\partial \over \partial x_i} \right)
           G^R(x_j)
       + G^R \left( x_j, \partial^\mu J_{D\mu } \right) = 0 \;  ,
\end{equation}
where $G^R(x_i)$ is a renormalized Green's function with $n_A$ external gluon
lines, $n_\psi$ quark (antiquark) lines, and $n_\omega$ ghost (antighost)
lines. $G^R(x_i, \partial^\mu J_{D\mu })$ is the same Green's function
inserted with divergence of the dilatation current. In momentum
space, we have,
\begin{equation}
         \left[-\sum_i p_i {\partial\over \partial p_i} -4(n_A + n_\psi +
     n_\omega-1) +(n_A + {3\over 2} n_\psi + n_\omega)\right] G^R(p_i)
       + G^R(p_i, \partial^\mu J_{D\mu }) = 0\;  .
\end{equation}
Here momenta $p_i$ are conjugation of all $x_i$ except for
one which is taken to be zero. On the other hand, a simple dimensional
analysis yields,
\begin{equation}
         \left[\sum_i p_i{\partial\over \partial p_i}
                      + \mu {\partial \over \partial \mu}
          + m^R{\partial \over \partial m^R} + 4(n_A + n_\psi +
     n_\omega-1)
           - \Big(n_A + {3\over 2} n_\psi + n_\omega \Big)\right]G^R(p_i)=0
                      \;  ,
\end{equation}
where $\mu$ is a renormalization scale.
Combining the above two equations, we get,
\begin{equation}
        \left[\mu{\partial \over \partial \mu} + m^R{\partial\over \partial
m^R}
   \right]G^R(p_i) ++ G^R(p_i, \partial^\mu J_{D\mu }) =0 \;  .
\end{equation}
According to Ref.~\cite{CDJ}, an insertion of the divergence of the
dilatation current can be replaced by derivatives with respect to the
gauge coupling, quark masses, and the gauge-fixing parameter.  Thus,
we find the following anomalous ward identity (Callan-Symanzik
equation),
\begin{equation}
         \left[\mu {\partial \over \partial \mu}
    + \beta {\partial \over \partial g^R}
    - \gamma_m m^R{\partial \over \partial m^R}
    -\gamma_3 \xi^R {\partial \over \partial \xi^R}
    +n_A{\gamma_3\over 2}+n_\psi{\gamma_2\over 2}+n_\omega{\tilde
 \gamma\over 2} \right]G^R(p_i)=0 \;  .
\end{equation}
This is just the well-known renormalization
group equation which can be derived independently
by studying the dependence of the
renormalized Green's function on the renormalization
scale.

\section{Breakup of the Hadron Masses}

Let us first recapitulate the main results obtained in the
last section. First of all, we have shown explicitly
that the QCD energy momentum tensor
$T^{\mu\nu}$ (with vacuum subtraction)
is a finite composite operator and thus
a physical observable. Second, the tensor can be separated
uniquely into the traceless and trace parts, each of which
is a finite and scale-independent composite operator,
\begin{equation}
     T^{\mu\nu} = \bar T^{\mu\nu} + \hat T^{\mu\nu}\;  .
\end{equation}
So the separation at this level is completely physical.
Finally, in physical states the traceless part of the tensor
is effectively a sum of the renormalized, gauge-invariant quark
and gluon contributions,
\begin{equation}
     \bar T^{\mu\nu} = \bar T^{\mu\nu R}_q(\mu^2)
     + \bar T^{\mu\nu R}_g(\mu^2) \;  ,
\label{lsep}
\end{equation}
and the trace part is a sum of the gauge-invariant quark mass
and trace anomaly contributions,
\begin{equation}
   \hat T^{\mu\nu} = \hat T^{\mu\nu R}_a(\mu^2)
                      + \hat T^{\mu\nu R}_m(\mu^2) \;  .
\label{tsep}
\end{equation}
The separation at this second level is renormalization and regularization
scheme dependent.

In this section, we study a breakup of hadron masses according to
the above properties of the energy-momentum tensor. We first discuss
matrix elements of various parts of the tensor in hadron
states. Then we write down the finite form of the QCD hamiltonian and
drive a mass separation formula. We revisit its application to the
nucleon and discuss implications of the result on the nucleon's
quark-gluon substructure and on the low-energy dynamics
of QCD\@. Application to the pion is studied in a
separated subsection, where we show that the color electric and
magnetic fields in the pion is the same as those in the QCD vacuum in
chiral limit. We also discuss the color fields in the QCD vacuum in a
Lorentz covariant, non-perturbative regularization scheme.

\subsection{ Matrix Element of the Energy-Momentum Tensor in Hadrons}

Let us consider the forward matrix element of the energy-momentum
tensor in a hadron state, $|P\rangle$, where $P^\mu$ is the four-momentum
of the state. We assume the state is normalized according  to
$\langle P|P \rangle = (E/M)(2\pi)^3\delta^3(\bf{0})$, where $E=P^0$ is the
energy of the state and $M$ is the mass of the hadron. It is well known
that,
\begin{equation}
          \langle P | T^{\mu\nu}| P \rangle =P^\mu P^\nu/M \;  .
\label{tmatrix}
\end{equation}
A simple derivation of the above equation goes like this: According
to Lorentz symmetry, one has,
\begin{equation}
          \langle P | T^{\mu\nu} | P \rangle =a P^\mu P^\nu + bg^{\mu\nu} \;  .
\label{gmatrix}
\end{equation}
where $a$ and $b$ are scalar constants. On the other hand, the hamiltonian
of the system is $ H = \int d^3 \vec{x}\  T^{00}$, which has the following
matrix element in the hadron state,
\begin{equation}
         \langle P |H| P \rangle = (E^2/M) (2\pi)^3
         \delta^3(\bf{0}) \;  .
\label{hmatrix}
\end{equation}
Comparing Eq.~(\ref{hmatrix}) with Eq.~(\ref{gmatrix}), we
obtain Eq.~(\ref{tmatrix}).

Eq.~(\ref{tmatrix}) allows us to obtain the matrix elements of the
traceless and trace parts of the energy-momentum tensor separately.
In fact, decomposing both sides of Eq.~(\ref{tmatrix}) into trace
and traceless parts, we have,
\begin{eqnarray}
\langle P| \bar T^{\mu\nu} |P \rangle
           & = &  \left( P^\mu P^\nu - {1\over 4}g^{\mu\nu} M^2 \right) /M
                      \;  ,  \nonumber \\
\langle P| \hat T^{\mu\nu} |P \rangle
           & =&  {1\over 4}g^{\mu\nu} M \;  .
\label{smatrix}
\end{eqnarray}
The right-hand sides of both equations are independent of the
renormalization scale, as required by Lorentz symmetry. The
second line in (\ref{smatrix}) is not new; it has appeared
before in the literature~\cite{JM,SVZ,CHEN}.

We use Lorentz symmetry again to define two scheme-dependent matrix
elements below. First, we define matrix element
of the quark operator appearing in the traceless tensor,
\begin{equation}
         \langle P| \bar T^{\mu\nu}_q |P \rangle  =
        a(\mu^2) \left(P^\mu P^\nu - {1\over 4}g^{\mu\nu} M^2 \right) /M
                      \;  ,
\end{equation}
where $a(\mu^2) $ has an explicit scale-dependence. From Eq.~(\ref{lsep})
and the first line in Eq.~(\ref{smatrix}), we have,
\begin{equation}
         \langle P| \bar T^{\mu\nu}_g |P \rangle =
        \left(1-a(\mu^2) \right)
           \left( P^\mu P^\nu - {1\over 4}g^{\mu\nu} M^2 \right) / M \;  .
\end{equation}
Second, we define matrix element of the quark-mass operator,
\begin{equation}
      \langle P | \hat T^{\mu\nu}_m |P \rangle
     = b(\mu^2)  {1\over 4}g^{\mu\nu} M \;   ,
\end{equation}
where the renormalization scale-dependence comes entirely from
the anomalous dimension $\gamma_m$ of the mass operator, which depends on the
renormalized gauge coupling. From Eq.~(\ref{tsep})
and the second line in (\ref{smatrix}), we have,
\begin{equation}
      \langle P | \hat T^{\mu\nu}_a |P \rangle
     = \left(1-b(\mu^2) \right)  {1\over 4}g^{\mu\nu} M \;  .
\end{equation}
Thus the matrix elements of the four parts of the energy-momentum tensor
are entirely determined by two parameters $a(\mu^2)$ and $b(\mu^2)$.

The matrix element $a(\mu^2)$ can be extracted from hadron
structure functions measured in deep
inelastic scattering. According to operator product expansion,
the twist-two operator $\bar T^{\mu\nu}_q$ appears
in the expansion of product of two vector
or axial-vector currents. Using a dispersion relation,
one can relate the matrix element of $\bar T^{\mu\nu}_q$
to the first moment of quark distributions,
\begin{equation}
        a(\mu^2) =
             \sum_f \int^1_0 dx \, x
           \left[ q_f(x,\mu^2)+\bar q_f(x, \mu^2) \right]
\end{equation}
where $q_f$ and $\bar q_f$ are quark and antiquark distributions
in the hadron. The physical meaning of $a(\mu^2)$ is the momentum
fraction of the hadron carried
by quarks in the infinite momentum frame~\cite{YND}.

The matrix element $b(\mu^2)$ is proportional to
the nucleon's scalar charge $\langle P |\bar \psi\psi |P \rangle$,
which cannot be measured directly from an experiment.
However, the operator
$\bar\psi m \psi$ is a part of the QCD lagrangian which
breaks the chiral symmetry explicitly. Therefore, useful
information about $b(\mu^2)$
may be obtained by studying physical effects of chiral symmetry
breaking.

\subsection{A Separation of Hadron Masses}

The QCD hamiltonian is defined as,
\begin{equation}
   H_{\rm QCD} = \int d^3\vec{x} \, T^{00} (0,\vec{x}) .
\end{equation}
{}From the discussion of the last section, $H_{\rm QCD}$
is a finite operator. [Interestingly, however, the lagrangian density
has $1/\epsilon$ divergences.] According to the separation
of the energy-momentum tensor, we have the partition
of the hamiltonian operator,
\begin{equation}
  H_{\rm QCD} = H_q' + H_g + H_m' + H_a  \;  ,
\label{ham}
\end{equation}
where various terms are,
\begin{eqnarray}
H_q' &=&  \int d^3\vec{x}  \left[\psi^\dagger(-i{\bf D\cdot \alpha})
           \psi + {3\over 4}\bar \psi m \psi\right] \;  ,
           \nonumber \\
H_g & =&  \int d^3\vec{x} \, {1\over 2}({\bf E}^2+{\bf B^2}) \;  ,
           \nonumber \\
H_m' &= &  \int d^3\vec{x} \, {1\over 4}(1+\gamma_m)\bar \psi m\psi  \;  ,
           \nonumber \\
H_a &=&  \int d^3\vec{x} \, {1\over 4}\beta(g)({\bf E}^2-{\bf B^2}) \;  .
\label{sham1}
\end{eqnarray}
Rearranging the mass term by defining,
\begin{eqnarray}
H_q  &=&  \int d^3\vec{x} \, \psi^\dagger(-i{\bf D\cdot \alpha})
           \psi  \;  , \nonumber \\
H_m &=&  \int d^3\vec{x}
           \left(1+{1\over 4}\gamma_m \right)
           \bar \psi m\psi  \;  ,
\label{sham2}
\end{eqnarray}
we have Eq.~(\ref{ham}) without the primes. Note that the above expression
implicitly contains the renormalization counter terms, which
we omit for simplicity. The physical meaning of
the various pieces of the hamiltonian is clear:
$H_q$ represents the quark kinetic plus potential energy (the static
color interaction is not included); $H_g$ is the gluon kinetic
and potential energy; $H_m$ is the quark mass contribution; and
finally, $H_a$ is the anomaly contribution whose significance
will become clear later.

In field theory, masses of bound states are usually defined as poles
in Green's functions (implicitly so in lattice calculations).
In hamiltonian formalism, one can define
masses as eigenvalues of a hamiltonian as in non-relativistic
quantum mechanics. In the following discussion,
we assume that hadron masses are calculated as expectation values
of the hamiltonian operator in the rest frame of hadrons,
\begin{equation}
      M = {\langle P |H_{\rm QCD}|P \rangle \over
      \langle P| P\rangle}|_{\rm rest~frame} \;  .
\label{defmass}
\end{equation}
We view the above equation as a mass probe into the structure of quark
and gluon configurations in hadron states, since different pieces of
the hamiltonian have different sensitivity to various components of
hadron wave functions.  At this point, the reader might recall the
recent investigation in the literature about the spin structure of the
nucleon~\cite{JM}, where one is interested in how the spin of the
nucleon is made of the spin and orbital angular momenta of quarks and
gluons.  The same question can be asked of the mass structure of
hadrons.

According to Eqs. (\ref{sham1}), (\ref{sham2}), and (\ref{defmass}),
masses of hadrons can be separated into various contributions,
\begin{equation}
     M = M_q + M_g + M_m + M_a \;  .
\end{equation}
Relating the matrix elements of different pieces of the hamiltonian
to those of the energy-momentum tensor, we have,
\begin{eqnarray}
    M_q &=&  {3\over 4}
           \left( a-{b\over 1+\gamma_m} \right) M  \;  ,
           \nonumber \\
    M_g &=&  {3\over 4}(1-a)M  \;  ,
           \nonumber \\
    M_m &=&  {4+\gamma_m\over 4 ( 1+\gamma_m)}b M  \;  ,
           \nonumber \\
       M_a &=&  {1\over 4}(1-b) M \;  .
\end{eqnarray}
Therefore, knowing the parameters $a$ and $b$, we can determine
a complete separation of the masses. In the following subsections,
we shall apply the separation to several cases and discuss physical
significance of the result.

\subsection{The Mass Structure of the Nucleon}

The mass structure of the nucleon was analyzed in detail in the Letter
paper~\cite{JI}. Let us briefly summarize the main result here
and discuss its physical implications more thoroughly. In the
end, we shall make some remarks for the mass separation of other hadrons
following a rather plausible assumption.

The matrix element $a(\mu^2)$ for the nucleon has been measured quite
accurately in various high-energy scattering processes involving nucleons.
A recent extraction gives~\cite{CTEQ},
\begin{equation}
   a_{\rm \overline{MS}}(1 \hbox{ GeV}^2) =0.55 \ .
\end{equation}
In the Letter, two estimates of $b(1 \hbox{ GeV}^2)$,
in the limits of chiral SU(3) ($m_s\to 0$) and heavy strange quark
($m_s\to\infty$), were given. They are,
\begin{eqnarray}
      b(m_s\to 0)  &=&  0.17   \;  , \nonumber \\
      b(m_s\to\infty) &=&  0.11 \;  .
\end{eqnarray}
Since the two limits do not lead to qualitatively different
conclusions, we focus on the chiral limit below.
The numerical numbers in this case are,
\begin{eqnarray}
       M_q  &=&  270 \hbox{ MeV} \;  , \nonumber \\
       M_m  &=&  160 \hbox{ MeV} \;  , \nonumber \\
       M_g  &=&  320 \hbox{ MeV} \;  , \nonumber \\
       M_a  &=&  190 \hbox{ MeV}\;  .
\end{eqnarray}

According to the above result, the quark kinetic and potential
energies contribute about 1/3 of the nucleon mass. The further
separation into the two is not gauge invariant. However,
the practice does have an intuitive appealing at the phenomenological
level. If there are three massless quarks confined to a spherical
cavity of radius 1 fm, as in the MIT bag model,
the total kinetic energy is
$600 \sim 700$ MeV\@. Thus, the color-current interaction
between quarks and gauge fields contributes
$-300 \sim- 400$ MeV to the mass, which is consistent
with the magnitude of $N-\Delta$ splitting.
Such large current interaction is intrinsic
to a truly relativistic theory. It does not occur
for instance in low-energy QED, where the static
Coulomb potential plays a dominant role. The strong current interaction
certainly induces quark interactions of Nambu-Jona-Lasino type, and
is perhaps at the origin of the chiral symmetry breaking.

The quark energy can be further separated into contributions of different
flavors. The parameter $a$ measures the fraction of the nucleon
momentum carried by quarks, which is known separately for
each flavor. For instance, 0.38 fraction of the nucleon momentum is carried
by up quarks, which can be translated into the up quark contribution
to the nucleon mass 250 MeV\@. Likewise, down quarks contribute
105 MeV mass, and strange quark pairs contribute $-85$ MeV\@.
One might try to break the contribution from each flavor into these
from valence and sea quarks. However, since the valence and sea
contributions to the matrix element $b$ are unknown,
the separation is not possible. Nonetheless, for up and down quarks,
the contribution from $b$ is not large, and may be neglected.
{}From the momentum fractions carried by the sea, we find that the
up or down sea contribution to the nucleon mass is on the level of 30 MeV\@.
The small number indicates that there are not many quark and antiquark
pairs in the wave function. Thus the nucleon seems to have a simpler
Fock expansion than the small current quark masses imply.

The quark mass term contributes about 1/8 of the nucleon mass.
About half of which comes from the strange quark pairs.
The strange quark contribution here is definitely less certain than
two light flavors. One might hope that future lattice measurement
of the strange scalar charge may reduce the uncertainty.
The implied magnitude of the scalar charges $\langle P|\bar uu|P\rangle$,
$\langle P|\bar dd|P\rangle$,
which are charge conjugation odd quantities, is another indication
that the number of quark-antiquark pairs is small.

The normal gluon energy contributes about 1/3 of the nucleon mass.
This energy includes the color-static Coulomb energy between quarks.
The gluon part of the trace anomaly contributes about 1/4 of the mass.
{}From these, we deduce the color-electric and color-magnetic
fields in the nucleon separately (taking $\alpha_s(1 \hbox{ GeV}^2)=0.4$),
\begin{eqnarray}
     \langle P|{\bf E}^2|P\rangle &=& 1700 \hbox{ MeV}  \;  ,
           \nonumber \\
     \langle P|{\bf B}^2|P\rangle  &=&  -1050 \hbox{ MeV} \;  .
\end{eqnarray}
The second line indicates that the color magnetic field in the nucleon
is smaller than that in the vacuum.
This property of the magnetic field has long been suspected
phenomenologically. The present result lends a strong support
for the educated guess. Clearly, this behavior
of color fields is closely related to color confinement.

In the chiral limit,
the trace anomaly contribution
is analogous to the vacuum energy
in the MIT bag model. In fact, the trace part of the
energy-momentum tensor in the bag is $Bg^{\mu\nu}$, where
$B$ is the energy density of the ``perturbative vacuum.''
The role of such energy density is to confine quarks. Thus
the scale symmetry breaking is explicitly connected to
quark confinement. It is essential then to include the
effects of the trace anomaly in phenomenological
hadron models.

A final comment
is about the role of strange quarks. They
contribute $-60$~MeV to the mass through the trace anomaly
because the $\beta$ function depends on the number of flavors.
The kinetic and potential
energy contributes about $-85$~MeV\@. Adding these to the
strange mass contribution of 115~MeV, one gets a total
of $-30$~MeV, roughly three percent of the total mass.
Therefore the uncertainty in the separation is
largely limited to the strange sector.

What shall be the general feature of the
mass separation for other hadrons for which
there are no data?
First of all, so long as one is concerned with non-strange
hadrons, the contribution
of the quark mass term is presumably small and we may neglect $b$.
Second, that gluons
carry about half of the nucleon momentum in infinite
momentum frame is perhaps approximately
true for all hadrons, e.g., $\rho$ or $\Delta$. Thus we further assume
$a=0.5$. Given these guesses, we have,
\begin{equation}
           M_q = {3\over 8} M \; ,
           \qquad   M_g = {3\over 8} M \; ,
           \qquad    M_a = {1\over 4} M  \; .
\end{equation}
This is a heuristic way to sum up the main result of the mass
separation.

\subsection{The Mass Structure of the Pion and the Vacuum}

The mass structure of the pion is particularly interesting because,
according to the Goldstone theorem, the pion is intrinsically
different from ordinary hadrons: it is
a collective mode in the QCD vacuum. As we shall see, the mass
structure indeed reflects this.

The matrix element $a(\mu^2)$ can be extracted from
the quark distributions measured in
$\pi-N$ Drell-Yan processes. The quality of the available data~\cite{PDY},
however, is much less satisfactory compared with that of the quark
distributions in the nucleon. Nonetheless, it seems safe to conclude
that $a(1 \hbox{ GeV}^2)$ is known at ten percent level, with a
central value similar to that of the nucleon,
\begin{equation}
         a(1 \hbox{ GeV}^2) = 0.55 \pm 0.05 \;  .
\end{equation}
[Note that the precision of the data is not good enough to discern
radiative corrections at the sub-leading-logarithmic level, so
we have suppressed the scheme label.]
Thus as far as high energy probes are concerned, the pion
is not dramatically different from other hadrons. This
is also true for the $\pi-N$ total cross section
at high energy, for which the quark counting rule appears
valid.

The matrix element $b$ can be calculated through study of
the pion mass in chiral perturbation theory. To avoid kinematic
singularity in chiral limit, we adopt the normalization
of the $\pi$ state, $\langle P|P\rangle = 2E(2\pi)^3\delta^3(\bf{0})$.
Therefore the matrix element $b$ is,
\begin{equation}
         bm_\pi^2 = \left\langle P|\sum_{f=u,d,s}
                m_f \bar \psi_f\psi_f |P \right\rangle \;  .
\label{defb}
\end{equation}
On other other hand, the first-order chiral perturbation
theory predicts~\cite{YND},
\begin{equation}
         m_\pi^2 = -(m_u +m_d)\langle 0|\bar uu + \bar dd|0\rangle/f_\pi^2
                      \;  .
\end{equation}
where $f_\pi$ is the pion decay constant and $|0\rangle$
is the QCD vacuum. The above equation tells us two things.
First strange quarks do not contribute
to the pion mass in the first-order perturbation, and
matrix element $\langle P|\bar ss|P\rangle $ is strongly suppressed.
Second, $m_\pi^2 \sim m_u, m_d$, and so to the first order accuracy,
one can take the chiral limit of the pion wave function in
the right hand side of Eq.~(\ref{defb}).

The pion mass can also be calculated by using the ordinary first-order
perturbation theory~\cite{GL},
\begin{equation}
        {1\over 2}m_\pi^2 = \left( P|m_u\bar uu + m_d\bar dd|P \right)  \;  ,
\end{equation}
where $|P)$ is the pion wave function in chiral limit. Comparing
the above equation with Eq.~(\ref{defb}), we have,
\begin{equation}
       b = 1/2 \;  .
\end{equation}
So the first-order chiral perturbation gives a clean prediction.

Using the above matrix elements, we roughly have,
\begin{equation}
       M_q  = 0  \; ,  \qquad
       M_g  = {3\over 8} m_\pi     \; ,  \qquad
       M_m  = {1\over 2} m_\pi    \; ,  \qquad
       M_a  = {1\over 8} m_\pi   \; .
\end{equation}
Two comments can be made immediately with regard to the
above mass partition. First, the quark kinetic and potential energies
cancel almost exactly. This fact is difficult to
reproduce in quark models for the pion, where quarks
carry large kinetic energy when confined to a small region
of space. Second, the color electric and magnetic
fields in the pion approach those in the vacuum in chiral
limit. This strongly indicates that the pion is a
collective excitation of the QCD vacuum.

It is tempting to use the above formalism to study
the color electric and magnetic fields in the QCD vacuum.
However, the energy-momentum tensor without vacuum subtraction
(``normal-ordering'') is not finite and it is dangerous to
work with divergent
quantities. Nonetheless, if one finds a Lorentz covariant
and non-perturbative regularization scheme to define
the energy-momentum tensor without the subtraction,
one can make statements about
the vacuum color fields, except, of course,
discussion is regularization scheme-dependent
and is meaningful only in a carefully defined context.
[Perturbative regularization schemes are usually plagued with
the infrared renormalon problem, and hence are not useful
in this context~\cite{JI2}.]

For simplicity, let us neglect quarks and study pure non-Abelian
gauge theory. Since the vacuum is not characterized by
any four-vector, the matrix element of the traceless part of
the energy momentum tensor vanishes. That is,
\begin{equation}
              \langle 0|\bar T^{\mu\nu}|0\rangle =0 \;  .
\end{equation}
Taking $\mu=\nu=0$, one gets,
\begin{equation}
              \langle 0 |{\bf E}^2 |0 \rangle = -
       \langle 0 |{\bf B}^2 |0 \rangle \;  .
\end{equation}
The color electric field is negative of the color magnetic field
in vacuum!
This appears to be a quite dramatic statement at first. However,
it is trivially true, with some caveats, in lattice QCD calculation .
On lattice, the color electric and magnetic fields can be defined as
the average of the trace of elementary plaquettes in space-time
and space-space planes.
Due to hypercubic symmetry on lattice, the two different plaquettes
have the same expectation value in vacuum. Remembering that the
electric field in  Minkowski space is related to that in Euclidean
space by a factor of $i$, the imaginary unit, one gets the above
equation immediately.

The trace part of the energy-momentum tensor contains
only the anomaly. According to Lorentz symmetry,
\begin{equation}
   \langle 0 | \hat T^{\mu\nu} |0 \rangle = g^{\mu\nu} \rho
\end{equation}
where $\rho$ is the vacuum energy-density in a particular
regularization scheme. Using the trace
anomaly, one has,
\begin{equation}
    \left\langle 0|(-2\beta/g^R) F^{\alpha\beta}F_{\alpha\beta}|0
           \right\rangle
   = 4\rho
\end{equation}
The above equation shows that the vacuum energy density is related
to the expectation value of $F^2$, which is called the gluon
condensate in the literature. Knowing the condensate in a
particular scheme, one obtains the color electric and magnetic fields of
the vacuum, and hence the vacuum energy density.

Unfortunately, no one has yet proposed a natural Lorentz covariant,
non-perturbative regularization scheme for the unsubtracted
energy-momentum tensor operator. However, there is a phenomenological
definition of the vacuum condensate used in the QCD sum rule
calculation~\cite{SUM}.
The magnitude of the condensate has been determined through
fitting to hadron spectra,
\begin{equation}
      \left\langle 0|{\alpha_s\over \pi}F^2|0 \right\rangle
           = (0.35 \hbox{ GeV})^4 \;  .
\end{equation}
This translates to,
\begin{equation}
              \langle 0 |{\bf B}^2 |0 \rangle = -
       \langle 0 |{\bf E}^2 |0 \rangle = 3.68 \hbox{ GeV}/ {\rm fm}^3 \;  .
\end{equation}
It is a rather large number from the phenomenological point of view. However,
one should not forget its definition when it is used in a
physical context.

\section{Discussions and Comments}

The structure of the nucleon is a subject that has
been discussed for many years.
The non-relativistic quark model has the virtue that it
is simple and captures many important aspects of physics.
Unfortunately, to improve our understanding, we must solve
QCD in non-perturbative region. Although lattice QCD provides
an effective method to calculate many observables,
it provides little insight about the physics.

The measurement of the spin structure functions of the nucleon
is a milestone in motivating new explorations of the quark-gluon
structure of the nucleon. It points to our deficiency in
traditional modeling
of hadrons and to a need for unquenched lattice calculations. On
the other hand, it also urges a better physical description
for hadrons. In light of this, any rigorous
information about the nucleon properties is useful.

The mass separation of the nucleon is certainly one step
towards a better understanding the quark and gluon dynamics of the
nucleon. As was discussed, it has many implications about the
physics of gluons and quarks, and their interactions. What is
particularly interesting is the anomaly contribution. If the
reader is familar with the ``spin crisis,''  he/she might recall that the axial
anomaly was considered as one of the constributions to the nucleon's
spin~\cite{AR}. Unfortunately,
there the separation between the anomaly and normal contributions
are not quite clear. In particular, the question of gauge
invariance and factorization has not be solved satisfactorily.
Here the anomaly contribution to the nucleon mass
is unambiguous and its physics interpretation is quite
clear.

One may ask where one goes from here. First, one can confirm
the present result by doing lattice calculations.
One can measure different pieces of the hamiltonian in the
nucleon state. Such a calculation in the end may help to
test lattice approximations. Second, one can try to
build models which are consistent with the present mass
separation. One important conclusion here is that the anomaly
term must be added to models like Nambu-Jona-Lasino.
Without this term, it is difficult to include the confinement
effects.

\acknowledgements
I wish to thank J. Collins, R. Jackiw, R. L. Jaffe, K. Johnson,
and W. Wiese for useful conversations on the subject of this paper.

\end{document}